\documentclass{iau}
\usepackage{graphicx}

\title[IAUS291.~~UCXBs with high luminosity: a key for a new scenario] 
{Ultra-compact X-ray binaries with high luminosity: a key for a new scenario} 

\author[Konstantin Pavlovskii \& Natalia Ivanova]   
{Konstantin Pavlovskii\thanks{E-mails: {\tt pavlovsk@ualberta.ca; nata.ivanova@ualberta.ca} }
 \and  Natalia Ivanova} 
 
\affiliation{Dept. of Physics, University of Alberta, \\ 11322-89 Ave, Edmonton, AB, T6G 2E7, Canada }

\pubyear{2012}
\volume{291}  
\jname{\mbox{Neutron Stars and Pulsars: Challenges and Opportunities after 80 years}}
\editors{J. van Leeuwen, ed.} 
\begin{document}

\maketitle

\begin{abstract}
Ultra-compact X-ray binaries (UCXBs) are accreting systems with
periods less than 1 hour, which qualifies them to contain a
degenerate donor-companion. One would expect such systems to have the
easiest theoretical explanation, compared to other kinds of X-ray binaries.
Nonetheless, current theory fails to explain high mass transfer (MT) rates 
in three recently well observed long-period UCXBs. 
We find that this range of MT rates
can be maintained if the donor is a remnant of an out-of-thermal-equilibrium 
naked core of a giant which was
revealed in a very recent episode of a common envelope (CE) event.
\end{abstract}

\firstsection 
\section{Too high MT rates in UCXBs}

\begin{figure}[h]
\begin{center}
 \includegraphics[width=4.4in]{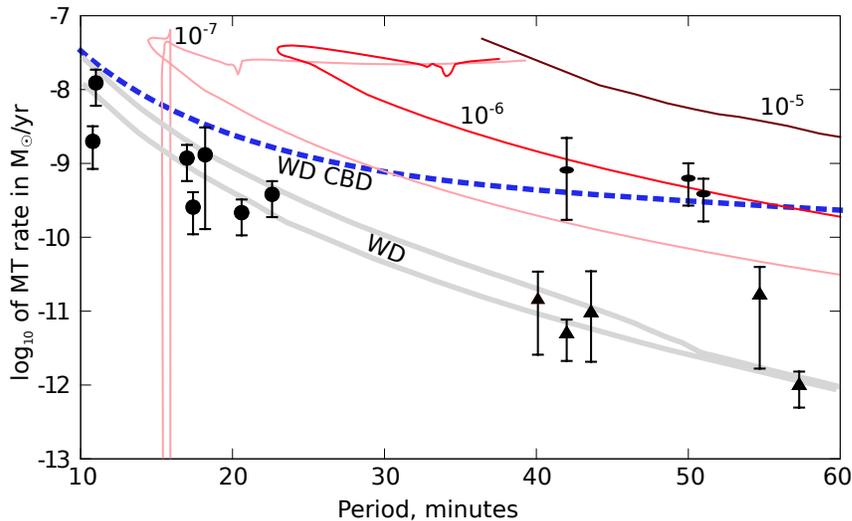} 
 \caption{Periods and MT rates for observed UCXBs compared to theoretical MT tracks of: 
a) partially degenerate WD donors (thick solid lines); 
b) WD donors evolved with a tidal torque of CBD where the mass ending up in CBD is adopted to be 
10 times higher than in a calibrated model (thick dashed line);
c) He remnants evolved under gravitational wave radiation after their formation with fast initial post-CE mass loss rate 
(thin lines, mass loss rates are indicated on the diagram in units of $\rm{M_{\odot}/yr}$). 
Observational data for persistent sources are shown with circles and ellipses 
(ellipses denote sources with anomalously high MT rates) and for transient sources with triangles (\cite{Heinke12}). 
He remnants were obtained using the stellar evolution code \texttt{MESA} (\cite{Paxton11}).}
\label{fig1}
\end{center}
\end{figure}

Recent observations show that,  at long orbital periods $\ge 40$ minutes,
 there are two  groups of UCXBs widely separated in their MT 
rates (\cite{Heinke12}). 
The first group is well known and consists of transient sources with very low average MT rates, $\sim 10^{-11} M_\odot$ per yr.
However, unexpectedly, the other group consists of permanent sources with average MT rates of at least two order of magnitude higher for the 
same orbital periods (Fig. \ref{fig1}). While in the first group an UCXB can belong to either the Galactic field, or be located in the direction of the bulge or in a globular cluster,  
in the second group all three UCXBs -- 4U 1626-67, 4U 0614+09, 4U 1916-053 -- belong only to the Galactic field.

The first group of long-period transient UCXBs is rather well understood in terms of MT sequence of cooling white dwarfs (WDs). The consideration that WDs are not completely degenerate at the start of the MT, and hence have some final entropy in the center, provides a range in possible MT rates for the same period. This effect is minimal for large periods, as MT was going on for already very long time and WDs are close to completion of  their cooling (\cite{Deloye11}, see also Fig.~1).

As for the second group of long-period persistent UCXBs, gravitational wave radiation can not provide such high MT rate even if a WD donor has finite entropy. 
Tidal torque provided by a circumbinary disk (CBD) can provide a stronger angular momentum loss leading to  higher MT rates. However, to explain the observed MT rates (as shown on Fig.~1), the fraction of mass ending in CBD has to exceed by 10 times the one predicted by the model calibrated on cataclysmic variables (\cite{Shao12}).
Other previously considered donors --  an initially slightly evolved main sequence donor with a He-rich core
 (\cite{1986ApJ...304..231N,2002ApJ...565.1107P}), or a naked He star produced by a CE  event (\cite{2008AstL...34..620Y}) -- 
will produce MT rates too low to match observations: 
at long orbital periods these donors were found to be almost fully degenerate.

\section{Alternative donors}

A naked He core that is formed in a CE event can be larger than a WD of the same mass.
It is crucial that if prior to the CE giant's core was non-degenerate, 
then during and immediately after the envelope 
ejection it experiences thermal readjustment. This leads to a fast 
MT onto a companion, with MT rates reaching $10^{-2}M_\odot{\rm yr^{-1}}$ at a peak.
After a peak, MT rates are smaller but still sufficient to keep the 
remnant out of thermal equilibrium for a while 
(\cite{2011ApJ...730...76I}).
How long this fast mass loss goes and how much of the core is removed during this mass loss, 
is not well established.

Hence, to understand high-MT UCXBs, we need to consider self-consistent He remnants.
While previous studies considered evolution of initially homogeneous He stars, 
we formed them via simplified CE event by evolving a $5~M_\odot$ star and stripping its hydrogen rich envelope.
We considered cases before the start of the He core burning prior a CE, and after.
The formed He remnants were evolved with different fast mass loss rates.
The MT rates as of UCXBs were then obtained assuming that these out-of-thermal-equilibrium 
remnants are placed in a binary with a neutron star companion to start the MT, and the 
binary evolution is driven by gravitational wave radiation only (see Fig.~1).
The remnants formed with an initially higher mass loss rate drive a higher MT 
rate under gravitational wave radiation because they are further out of their thermal equilibrium -- 
for the same remnant mass, they are more inflated and colder (see Figs.~2 and 3).

\begin{figure}[h]
\begin{center}
 \includegraphics[width=4.2in]{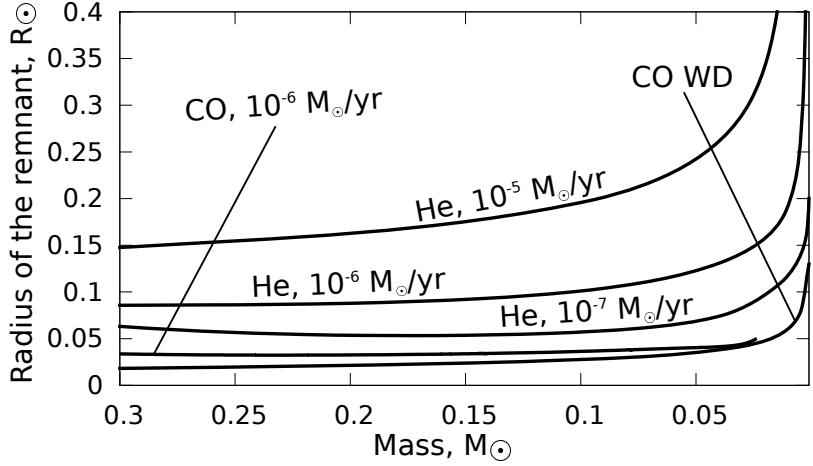} 
 \caption{Radii of He and CO remnants evolved with different mass loss rates.} 
\label{fig2}
\end{center}
\end{figure}

\begin{figure}[h]
\begin{center}
 \includegraphics[width=4.2in]{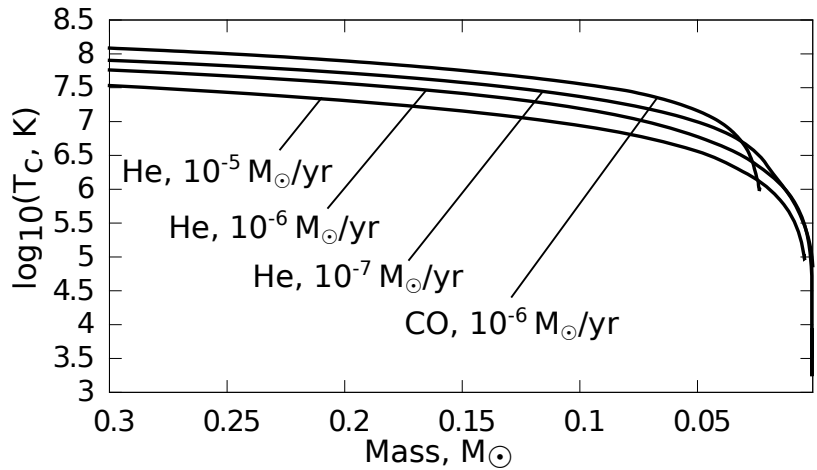} 
 \caption{Central temperatures of He and CO remnants evolved with different mass loss rate. } 
\label{fig3}
\end{center}
\end{figure}

We also find that the observed variations in He abundances in accretion disks of these mysterious fast-MT UCXBS do not necessarily require a WD donor. They can be explained by different durations of He core burning that took place in donor before or after the start of the MT. Depending on the post-CE orbital separation and on the mass and entropy of a He remnant, MT can start when the donor either:
\begin{itemize}
\item has not yet started He burning -- this can be applicable to 4U 1916-053, where the observed accretion disk is nitrogen-rich;
\item is going through He burning -- in this case the disk will be C/O rich, as observed for 4U 1626-67 and 4U 0614+09;
\item has already completed He burning. In this case  we find that the donor is less likely to be inflated enough to provide the observed MT rates.
\end{itemize}

\end{document}